\documentclass{vgtc}                          

\ifpdf
  \pdfoutput=1\relax                   
  \usepackage{graphicx}                
  \usepackage{gensymb}
  \usepackage{amsmath}
  \usepackage{tabularx}
  \usepackage{dblfloatfix}
  \usepackage{float}
  \DeclareGraphicsExtensions{.pdf,.png,.jpg,.jpeg} 
\else
  \usepackage{graphicx}                
  \DeclareGraphicsExtensions{.eps}     
\fi%

\graphicspath{{figures/}{pictures/}{images/}{./}} 
\usepackage[dvipsnames]{xcolor}
\usepackage{colortbl}
\usepackage{microtype}                 
\usepackage{authblk}
\PassOptionsToPackage{warn}{textcomp}  
\usepackage{textcomp}                  
\usepackage{mathptmx}                  
\usepackage{times}                     
\usepackage{cite}                      
\usepackage{tabu}                      
\usepackage{booktabs}                  
\usepackage[normalem]{ulem}

\definecolor{Red}{rgb}{1.0,0.0,0.0}

\newcommand{\prototypeone}[1]{PD2GAZE}
\newcommand{\prototypetwo}[1]{LED2GAZE}

\usepackage{todonotes}
\usepackage{enumitem}

\onlineid{1251}

\vgtccategory{Research}

\vgtcinsertpkg

\title{Optical Gaze Tracking with Spatially-Sparse Single-Pixel Detectors}

\newcommand*\samethanks[1][\value{footnote}]{\footnotemark[#1]}

\author[1,2]{Richard~Li\thanks{First and second authors contributed equally.}}
\author[1,2]{Eric~Whitmire\samethanks}
\author[2]{Michael~Stengel}
\author[2]{Ben~Boudaoud}
\author[2]{Jan~Kautz}
\author[2]{David~Luebke}
\author[1]{\\Shwetak~Patel}
\author[2,3]{Kaan~Ak\c{s}it}

\affil[1]{\scriptsize~University of Washington}
\affil[2]{\scriptsize~NVIDIA Research}
\affil[3]{\scriptsize~University College London}

\teaser{
\hspace*{-1cm}
 \includegraphics[width=1.1\textwidth]{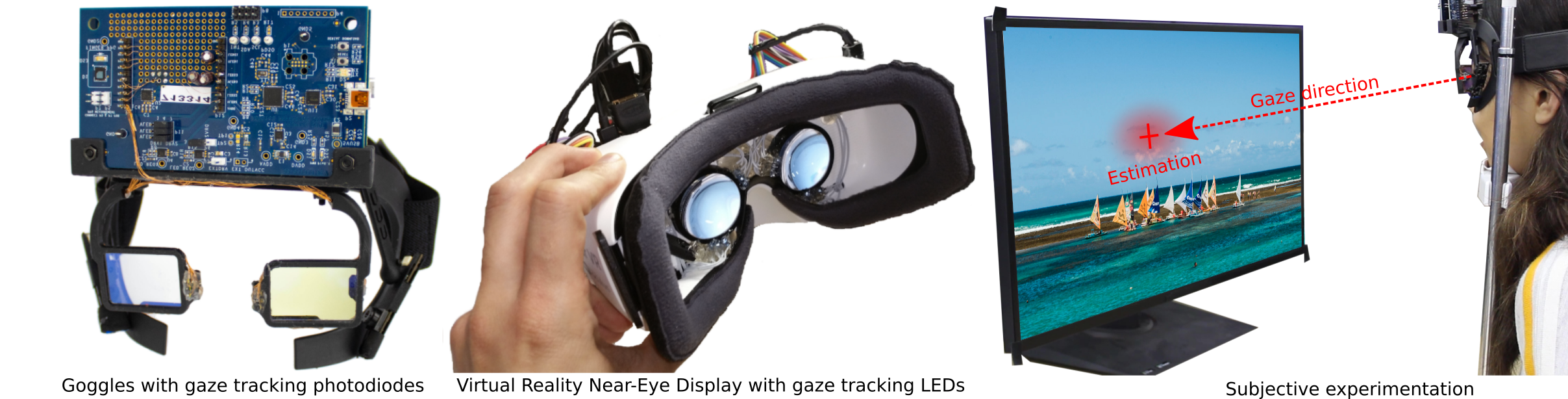}
 \caption{Gaze tracking, an essential component of next generation displays, needs to deliver several qualities such as accurate gaze estimation, low latency, small form-factor, low cost, low computational complexity, and a low power budget. To provide solutions for next generation displays, we demonstrate two single-pixel detector based gaze tracking prototypes. While the one shown on the left uses photodiodes and LEDs, the one shown in the middle uses only LEDs for both sensing and emitting light. As depicted on the right hand-side, we evaluate our gaze trackers with a series of subjective experiments.}
}

\abstract{
Gaze tracking is an essential component of next generation displays for virtual reality and augmented reality applications. Traditional camera-based gaze trackers used in next generation displays are known to be lacking in one or multiple of the following metrics: power consumption, cost, computational complexity, estimation accuracy, latency, and form-factor. We propose the use of discrete photodiodes and light-emitting diodes (LEDs) as an alternative to traditional camera-based gaze tracking approaches while taking all of these metrics into consideration.  We begin by developing a rendering-based simulation framework for understanding the relationship between light sources and a virtual model eyeball.  Findings from this framework are used for the placement of LEDs and photodiodes. Our first prototype uses a neural network to obtain an average error rate of $2.67\degree$ at $400$~Hz while demanding only $16$~mW.  By simplifying the implementation to using only LEDs, duplexed as light transceivers, and more minimal machine learning model, namely a light-weight supervised Gaussian process regression algorithm, we show that our second prototype is capable of an average error rate of $1.57\degree$ at $250$~Hz using $800$~mW.
} 

\CCScatlist{
  \CCScatTwelve{Human-centered computing}{Ubiquitous and mobile computing}{Ubiquitous and mobile devices}
  \CCScatTwelve{Computer systems organization}{Embedded and cyber-physical systems}{Sensors and actuators}
}

\begin{document}


\maketitle

\section{Introduction}
\label{section:Introduction}

Next generation displays~\cite{koulieris2019near} for virtual reality (VR) and augmented reality (AR) applications promise to improve our daily lives and routines. Gaze tracking is an essential and required component of these next generation displays, enhancing and enabling multiple methods and applications such as varifocal near-eye displays~\cite{dunn2017wide}, foveated near-eye displays~\cite{kim2019foveated}, super-resolution displays~\cite{akcsit2020patch}, and foveated computer graphics~\cite{tursun2019luminance}. 

While gaze tracking has largely remained a research tool, we believe that several factors have hindered the deployability of gaze tracking systems: accuracy, latency, power consumption, cost, computational complexity, and form-factor. Improvements in gaze tracking hardware and software in one of these metrics often involves compromising other metrics. However, for gaze tracking technology to enable applications in next generation displays, such technology has to lead to a useful quality gaze tracker with a small form-factor, low latency, and low power consumption.

In this paper, we explore means of designing a useful-quality gaze tracker that accounts for all of these metrics. Toward this end, we investigate techniques for simplifying both the hardware and software components of a gaze tracking system.
Concerned with the use of power and computationally demanding imaging sensors generally used for gaze tracking, we begin with a rendering-based simulation framework for exploring the possibility of decomposing cameras into individual single pixel sensors.
Using findings from these simulations, we place pairs of photodiodes and LEDs around a human subject’s eyes, modulating the light emission in a time multiplexed fashion and capturing the light reflected off of the eyes. We then use a fully connected neural network to process the recorded signals and estimate a gaze orientation, constituting our first standalone wearable gaze tracker prototype, NextGaze.
To further minimize cost and increase flexibility in manufacturing, we remove the need for photodiodes in NextGaze by taking advantage of the bidirectional characteristics of LEDs to enable receiving and emitting light with a single component, reducing the number of components used and the number of signals processed. We show that this second prototype, LED2Gaze, can reduce the accuracy error by up to half using a Gaussian process regression (GPR) model. Our contributions are listed as the following:

\begin{enumerate}[nolistsep]
    \item A rendering-based framework for simulating gaze tracking devices with arbitrary single-pixel sensors with new insights into the behavior of eye reflections.
    \item NextGaze, a wearable gaze tracking prototype equipped with photodiodes and LEDs. Our device obtains an average error of $2.67\degree$ at $400$~Hz while consuming only $16$~mW using a fully-connected neural network.
    \item LED2Gaze, a standalone gaze tracking prototype that uses LEDs both for emitting and sensing light. Our device obtains an average error of $1.57\degree$ at $250$~Hz consuming $800$~mW using a lightweight GPR model.
\end{enumerate}


 



\begin{table}[tp]
\caption{A comparison of six systems across four dimensions.  We position our work, the last two rows, as an unexplored middle ground across each of these dimensions.}
\begin{tabular}{lllll}
\hline
Name & Modality & Rate & Error & Power \\
\hline
Scleral Coil & Magnetic & 1000 Hz & 0.1° & 10+ W\\
SR Research & Camera & 1000 Hz & 0.33° & 24 W \\
Pupil Labs & Camera & 200 Hz & 0.6° & 1.5 mW \\
Tobii & Camera & 100 Hz & 0.5° & 900 mW \\
LiGaze & Photodiode & 128 Hz & 6.1° & 0.791 mW \\
CIDER & Photodiode & 4 Hz & 0.6° & 7 mW \\
\textbf{NextGaze} & \textbf{Photodiode} & \textbf{400 Hz} & \textbf{2.67°} & \textbf{16 mW} \\
\textbf{LED2Gaze} & \textbf{LED} & \textbf{250 Hz} & \textbf{1.57°} & \textbf{800 mW}
\end{tabular}
\label{tab:comparison_related_work}
\end{table}

\section{Related Work}
\label{section:RelatedWork}

We report and discuss the relevant literature on gaze tracking, and focus on three primary metrics: accuracy,  sample rate, and power consumption.

\subsection{Camera-Based Gaze Tracking}
\label{section:RelatedWork_CameraBased}

Video oculography is the most commonly used method for eye tracking.  Most video-based eye trackers rely on infrared illumination of the eye and an infrared-sensitive video camera that detects either the location of the pupil or glints on the cornea. A calibration procedure is used to construct a mapping between glint/pupil locations and gaze orientation.

A high end commercial video eye tracking system, such as the SR Research EyeLink 1000 Plus \cite{sr_research_2020}, is capable of sampling at 1000\,Hz with an average accuracy of $0.33\degree$.  More portable and affordable systems such as those produced by Tobii, SMI, and Pupil Labs operate at an order of magnitude lower sample rate, while maintaining a similar sub-degree accuracy.  However, the power consumption of these devices is generally on the order of multiple watts \cite{tobii_manual}.

In academic settings, the use of low resolution cameras for reducing the requirements of power consumption and computational resources needed for video oculography has seen promising results.  Borsato et al.~\cite{borsato2016episcleral} was able to significantly reduce processing and power requirements by simply repurposing the optical flow sensor of a computer mouse for tracking the episcleral surface (the white part) of the eye.  
While they were able to obtain an error bound of $2.1\degree$, the tracking was lost each time the user blinked, rendering it impractical for real-life use cases.  
Tonsen et al.~\cite{tonsen17_imwut} improved upon this accuracy by first simulating the different possible vantage points of cameras placed around a 3D model of the eye, eventually developing InvisibleEye, which leverages four millimeter-sized cameras of 5 x 5 pixels each to achieve a person-specific gaze estimation accuracy of $1.79\degree$.
Finally, Mayberry et al.~\cite{mayberry2016cider} developed CIDER to estimate the pupil location, but not the gaze orientation, hence we refrain from comparing their reported accuracy. However, we believe that CIDER was an exceptionally well-engineered system, with low power consumption (32 mW), high sampling rate (278 Hz), and detailed specifications given per component: camera, digitization, computation, and the near-infrared LEDs that were used for illumination.  

In this paper, we describe a simulation framework similar to the one used by Tonsen et al.~\cite{tonsen17_imwut} for exploring the possibility of removing the need for a camera altogether by using discrete single pixel detectors instead.
Ultimately, we use findings from these simulations to inform the design of a system that removes this significant component from CIDER's bill of materials, saving power, cost, and computation.

\subsection{Novel Sensing for Gaze Tracking}

In addition to traditional camera-based gaze tracking techniques, a number of novel sensing techniques have been explored to leverage other properties of the eyes for tracking.  On the high speed and invasive end, magnetic sensing has been employed to track scleral coils, wire coils embedded in a silicone ring that sits on the sclera of the eye~\cite{robinson1963method,collewijn1975precise}.  A voltage is induced in the wire coils when placed in a magnetic field, and that voltage is measured with thin physical connections to the coils \cite{robinson1963method}. This technique has been shown to offer a sampling rate of almost 10 kHz with an accuracy better than $0.1\degree$~\cite{collewijn1975precise}. This technique is generally used for lab-based studies and is not appropriate for consumer or mobile devices where power is a consideration.

At the other end of the spectrum, electrical sensing has been used to measure the voltage potential generated by the rotation of the eyes, which have an electric dipole between the cornea and retina, in a technique called electrooculography (EOG).  Although these signals are generally not robust enough for continuous gaze tracking, they can be used to detect movements such as blinks as well as relative direction of movement.  For example, Bulling and colleagues demonstrated wearable EOG goggles capable of detecting a set of eye gestures~\cite{bulling2009wearable}.  Furthermore, the eyeglasses company Jins produces a commercial product called the Meme with EOG electrodes embedded in the nose pad, which has been used to detect reading patterns~\cite{kai_reading}, to measure fatigue \cite{tag_alertness}, and to recognize facial expressions \cite{rostaminia2019w}.

Finally, our work falls under the category of spatially sparse optical sensing. While the cameras described previously use image sensors with many pixels, sparse optical sensing approaches position single-pixel optical sensors sparsely around the region of interest. Such approaches have the primary potential advantage of requiring fewer pixel sensors, eliminating the capture of pixels that would be otherwise redundant and resulting in lower dimensional data that requires less computational power and bandwith. In the case of head-mounted displays, sparse optical sensing also enables the possibility of moving the sensors out of the field of view for heightened immersion. For example, OLED-on-CMOS technology~\cite{oled_on_cmos} has demonstrated the ability to capture images with single-pixel sensors placed in-line and alternating with single-pixel displays. While there is a recent ongoing effort to make a gaze tracker product at industry using single-pixel sensors with scanning micro electromechanical systems (MEMS)~\cite{yang2019timer}, usage of single-pixel sensors for gaze tracking largely remains in research. For greater sparsity, Topal et al.~\cite{topal2014low} developed EyeTouch, which consisted of only eight IR LED and sensor pairs around the lenses of a pair of glasses, but required the use of a bite bar or other means of stabilizing the head.  LiGaze~\cite{li2017ultra} showed that it was possible to use a solar cell indoors for harvesting the power needed to perform eye tracking using photodiodes, achieving a sample rate of 128 Hz with $6.1\degree$~accuracy and consuming $791~\mu W$.  A sequel improved the power consumption to $395~\mu W$ with a slightly reduced sample rate of 120\,Hz \cite{li2018battery}.

\begin{figure*}[tp!]
	    \centering
	    \includegraphics[width=\textwidth]{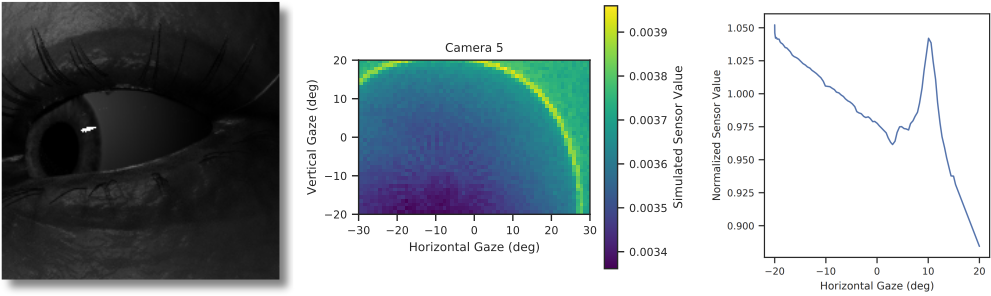}
	    \caption{Left: one camera's view of a simulated eye with corneal reflection.
	    Center: the simulated output of a single-pixel detector as a function of gaze angle, obtained by rotating the model shown in the left panel.
	    Right: real-life signal acquired by an actual photodiode as the eye rotated along the horizontal axis while staying constant vertically, verifying the distinct spike seen in the simulation shown in the center panel.  The signal in the right panel matches one row of the center panel (horizontal sweep, vertical constant).}
	    \label{fig:simulated_photosensor}
\end{figure*}

\begin{figure}[]
	\centering
	\includegraphics[width=1.0\columnwidth]{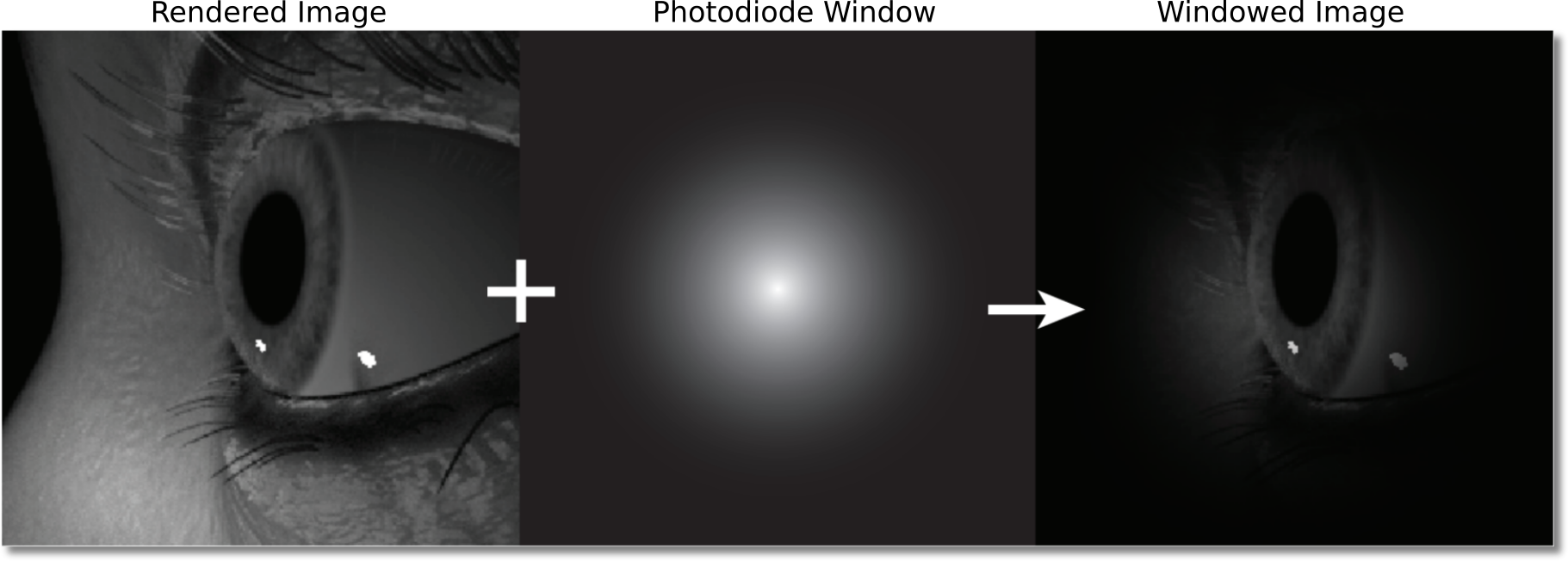}
	\caption{The simulation pipeline takes a rendered image from the point of view of a single-pixel detector, applies a windowing function to simulate the single-pixel detector lens, and sums the resulting image to simulate sensor output.}
	\label{nvgaze:fig:sim-window}
\end{figure}

SynthesEyes~\cite{wood2015rendering} and UnityEyes~\cite{wood2016learning} explored the possibility of using synthetic 3D models of human eyes for training models that can be applied to previously released, real-world datasets. Our work adopts a similar methodology of leveraging simulation findings for informing the design of our prototypes, with which we then implement and use to collect our own datasets for validation. Our functional prototypes gaze trackers are robust to head movement, removing the need for a bite bar, while improving the accuracy and sample rates given by LiGaze.  We position our work at the intersection of high performance (accuracy and sample rate) and low power consumption.

\section{Sparse Optical Gaze Tracking}
\label{section:SpatiallySparseOpticalGazeTracking}

The idea of sparse optical gaze tracking is to use single pixel emitters and receivers, spread out in physical space, to detect the gaze orientation of the user’s eye.
Depending on the orientation of the eye, the sensor will capture light directly reflected from the cornea (referred to as a "glint") and scattered from the iris, sclera, and skin.


\subsection{Rendering-Based Simulation Framework}

In order to inform the design of our sparse optical gaze tracking systems, namely the placement of the optical sensors and emitters, we constructed a framework to simulate how a gaze tracker would perform under different configurations. This framework uses a realistic 3D model of a human face with a parametrically-defined eyeball from the work by Kim et al.~\cite{kim2019nvgaze}. The eyeball can be rotated to any horizontal and vertical gaze direction and the pupil size can be adjusted from $2$~mm to $8$~mm. Moreover, the top and bottom eyelids can be adjusted from fully open to fully closed. The textures were adjusted to match the properties of skin under infrared illumination.

We place virtual cameras at the position and orientation of different proposed single-pixel sensor configurations following the guidance of Rigas et al.~\cite{rigas2018photosensor}. We use point light sources to simulate infrared emitters such as LEDs.
For a given facial configuration (eye gaze in $x$ and $y$, pupil size, eyelid position), we render an image from the viewpoint of each of the cameras. Because the sensitivity of a single-pixel sensor varies with the angle of incidence, we use a Gaussian distribution to represent sensitivity in accordance with the datasheet of a sensor. For each image, we transform all pixels in the image using the following Gaussian window function:

$$ g(x,y) = e^{\frac{-((x - x_{0})^2 + (y - y_{0})^2)}{2 \sigma ^ 2}},$$
where $x$ and $y$ represent the pixel coordinates, $x_0$ and $y_0$ represent the image centers, and $\sigma$ is the standard deviation of the Gaussian distribution that represents angular sensitivity of a single-pixel sensor. Figure \ref{nvgaze:fig:sim-window} summarizes the effect of this angular sensitivity of a single-pixel sensor. Finally, to simulate the accumulation of light at a single-pixel sensor, all pixels in the transformed image are summed as follows, where $i(x, y)$ represents a pixel from the original rendered image:

$$ s = \sum_{x}^{} \sum_{y}^{} i(x, y)*g(x, y),$$
An important observation from this process is the importance of using 16-bit rendering and taking care not to saturate the image. The images in Figure \ref{nvgaze:fig:sim-window} have been artificially brightened, for clarity, but note that the glints in these brightened images consist of saturated white pixels. If these images were used for the simulation, the signal due to the glint would be artificially weakened. To achieve a high-fidelity simulation, the simulated illumination must be decreased so that there are no clipped pixels from the direct reflections. Similar simulation techniques that either use rendered images \cite{zemblys2018making, rigas2017hybrid, rigas2018photosensor} or images captured from a camera \cite{irie2002laser} could be prone to this issue.

\begin{figure*}[t]
    \centering
    \includegraphics[width=1\textwidth]{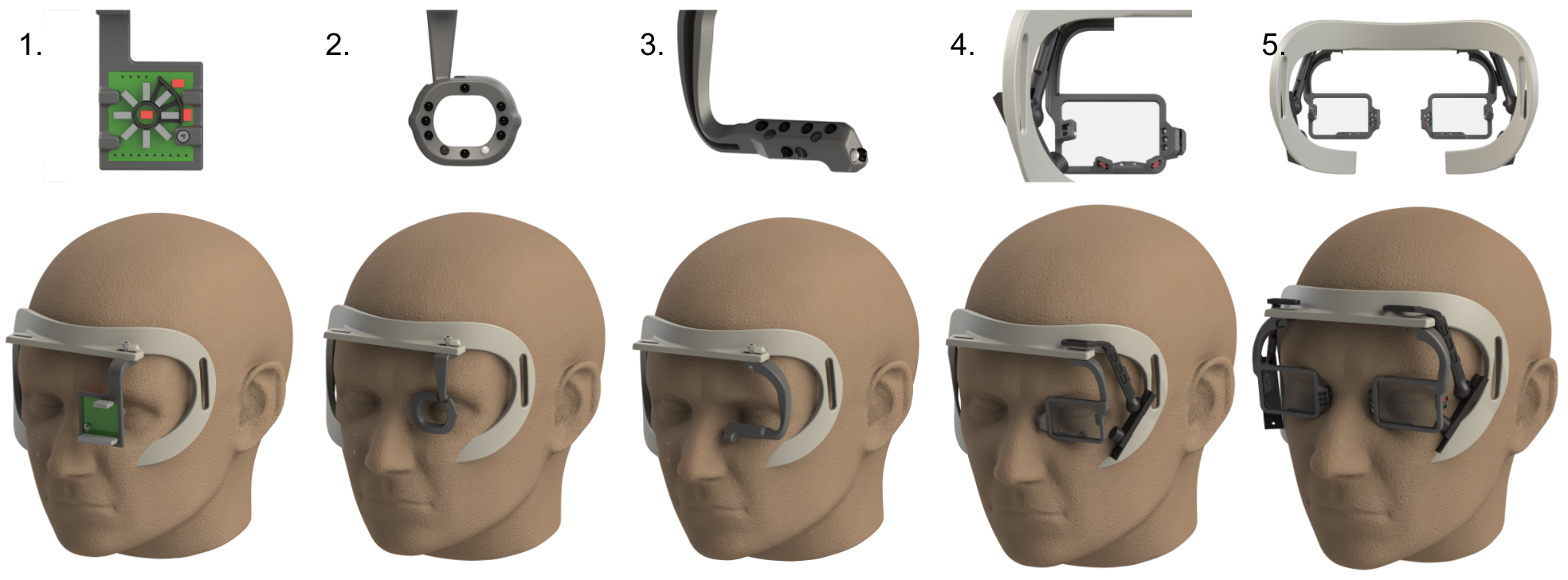}
    \caption{Examples of intermediate steps in our iterative co-design process between simulation and implementation.  From left to right:  1. Sensors placed directly in front of the eye, 2. sensors placed around the eye enabling looking through the lens, 3. increasing the field of view by experimentally moving all the sensors to the bottom, 4. adding a camera and some sensors along the vertical axis, and 5. streamlining the design.}
    \label{nvgaze:fig:nextgaze-iterative-design}
\end{figure*}

A second observation concerns the interaction of the glints with the edge of the cornea as the eye changes concavity. As the eye moves and a glint approaches the edge of the cornea, the glint becomes stretched and the received signal strength at the sensor increases as depicted in Figure~\ref{fig:simulated_photosensor}. The eye orientation at which this effect occurs depends on the position of both an emitter and a single-pixel sensor. Figure~\ref{fig:simulated_photosensor} shows the output of a single simulated single-pixel detector as a function of gaze directions, where the relatively smooth gradient along the lower left portion of the image is due to the pupil and glint moving within the sensitive region. The circular edge along the outsides of the image corresponds to gaze locations where the glint is positioned at the edge of the corneal bump. Within this bright edge, the glint is positioned on the cornea; outside of this edge, the glint is located on the sclera.

Center portion of Figure~\ref{fig:simulated_photosensor} shows an artificially brightened rendering corresponding to the gaze orientation of $27\degree$ horizontal and $-20\degree$ vertical. The smearing of the glint along the edge of the cornea causes more light to enter the single-pixel detector. Compare these simulated results to actual collected data as the eye pursues a target along a horizontal line from $-20\degree$ to $20\degree$ as shown in  Figure \ref{fig:simulated_photosensor} (right). We hypothesize that the spike around $10\degree$ corresponds to the glint aligning with the edge of the cornea. This effect has not been demonstrated or accounted for in prior work that relies on simulated data~\cite{zemblys2018making, rigas2017hybrid, rigas2018photosensor, katrychuk2019power}. A practical implication of this result is that it is best to minimize the number of point light sources active at a time. Multiple light sources will result in multiple glints that only complicate the tracking problem. Fewer light sources would maximize the smoothness of the transfer function. On the other hand, future work could consider using these direct reflections as a feature to model the geometry of the eye.

In addition to these two findings, a co-design process between simulated experiments and implementation iterations revealed that real life human faces varied so significantly that determining specific LED locations based on a simulation was not practical.
We show some of the intermediate designs in Figure \ref{nvgaze:fig:nextgaze-iterative-design}, and this iterative process helped reveal a number of design guidelines for working with spatially-sparse detectors:

\begin{itemize}
    \item As mentioned previously, minimize the number of emitters on simultaneously to avoid saturated receivers.
    \item Co-locating emitters and receivers generally provides the best signal.
    \item For maximum information gain with sparse sensors, the more diversity in perspective the better.
    \item However, staying below the eye is recommended to avoid eyelash interference.
\end{itemize}

We used these guidelines as we iterated through a number of designs, two of which we have selected to describe in this paper.

\begin{figure}[!h]
	\centering
	\includegraphics[width=0.8\columnwidth]{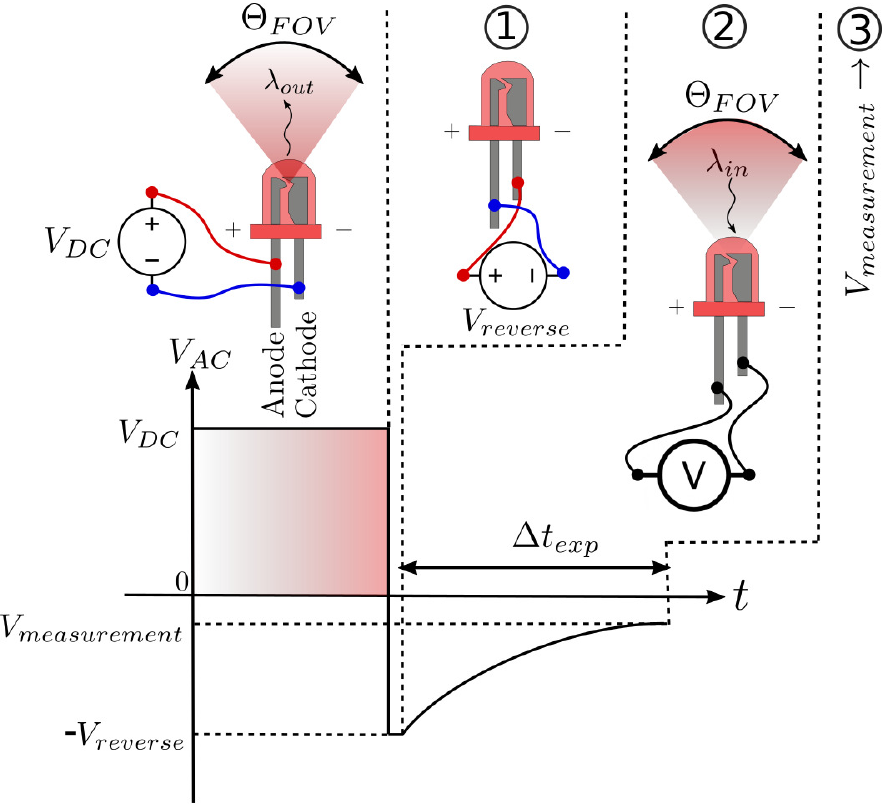}
	\caption{Different LED modes: (1) applying a forward voltage of $V_{DC}$, in which the LED emits light with a wavelength of $\lambda_{out}$ and an emission cone angle, $\Theta_{FOV}$; (2) applying a reverse voltage pulse, $V_{reverse}$, for a short time duration, discharging LED with incoming light that has a wavelength of $\lambda_{in}$ for a specific time, $\Delta t_{exp}$, with an reception cone angle of $\Theta_{FOV}$; and (3) measuring a voltage, $V_{measurement}$, from the LED.}
	\label{nvgaze:fig:reverse-led}
\end{figure}

\subsection{Single-Pixel Sensing Hardware}

\begin{figure*}[t]
    \centering
    \includegraphics[width=1\textwidth]{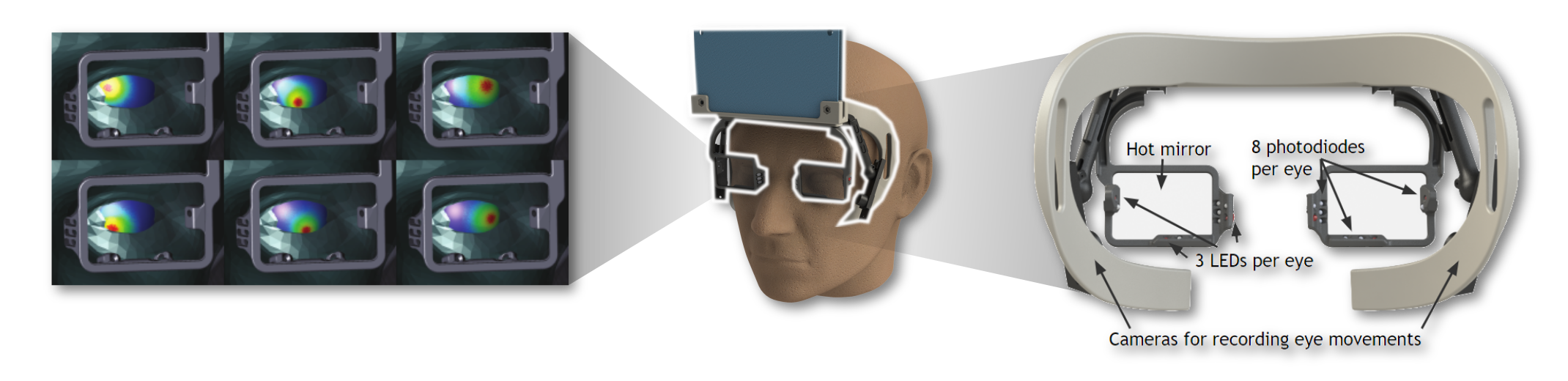}
    \caption{System architecture of NextGaze. Left: a Zemax optical simulation of six photodiode placements and their projected regions of sensitivity.  Right: the arrangement of optical and electrical components within the prototype, configured according to findings from the simulation.}
    \label{nvgaze:fig:nextgaze-rendering}
\end{figure*}

In order to leverage the findings from the simulation framework for our hardware implementation, we consider the use of LED and photodiode pairs and duplexed LEDs as single-pixel sensors. In the LED-photodiode pairs case, the LEDs are used to illuminate the eye and may be modulated to provide a more strategic image of the eye. LEDs with infrared light emission are typically used in gaze tracking hardware for NEDs, since humans are insensitive to IR illumination~\cite{dai20176}. A human eye's cornea has similar absorption and reflection characteristics in the near IR spectrum as in visible light~\cite{damianeye}. Furthermore, IR LEDs have a narrow bandwidth (typically around 50~nm), avoiding cross-talk with other wavelengths. The photodiodes are used to capture signals related to gaze direction.

In the latter case, we also propose the use of LEDs to both illuminate the eye and capture light. LEDs provide illumination when a forward voltage is applied to their two electrical terminals. However, LEDs can also act as photodetectors~\cite{dietz2003very}. This duplexing can be accomplished with three steps that are depicted in Figure~\ref{nvgaze:fig:reverse-led}.
Typically, LEDs are most sensitive to wavelengths $\lambda_{in}$ that are shorter than their emission spectrum (so $\lambda_{in} < \lambda_{out}$) \cite{lange2011multicolour}. Thus, larger exposure times are required if LEDs with the same emission spectrum are used. To achieve the lowest possible latency with a given configuration, we select different LEDs that have intersecting emission and sensing spectra in the IR range.

Eye safety is a very important aspect when a user is exposed to infrared radiation; $\Delta~t_{exp}$ and maximum irradiance of an LED must be considered according to safety regulations for infrared light sources . In our implementation, we followed a commonly accepted guideline~\cite{boucouvalas1996iec}, accordingly.

\section{Prototypes}
\label{section:Prototypes}

Based on learnings from the simulations and described sensing approaches, we developed our first prototype, NextGaze. Then, by simplifying both the software and hardware components, we arrived at our second prototype, LED2Gaze, which demonstrated improved accuracy while using fewer physical hardware components and a simplified estimation algorithm.

\subsection{Gaze tracking photodiodes}
Our first system consists of a ring of LEDs and photodiodes around each eye. The system is designed to be used as a standalone gaze tracker with an external display,
therefore it was constructed as part of a face mask. The mask attaches to the face using an elastic band. An external display is used to calibrate the gaze tracker prototype.

The full design is shown in Figure~\ref{nvgaze:fig:nextgaze-rendering}. The eye is illuminated by three LEDs embedded within the frame of the device. Two or three photodiodes are clustered around each LED. In total, there are eight photodiodes, placed strategically such that for a typical user, they will cover different parts of the eye. Figure~\ref{nvgaze:fig:nextgaze-rendering} shows the results of a Zemax optical simulation for a subset of six photodiodes. The images highlight the intended sensitive region for each photodiode.

To facilitate development and proper placement of the optical elements, a Pupil Labs infrared camera is placed just outside the user's field of view. A hot mirror in front of the eyes reflects infrared light from the eyes into the camera lens. 
The camera is used only for debugging purposes and is not part of the final sensing pipeline. No experiments were conducted using the camera since we can compare against the baselines already reported by camera-based gaze trackers as discussed in section~\ref{section:RelatedWork_CameraBased}, or specifically the specifications of the Pupil Labs camera~\cite{pupil_labs}.

The LEDs and photodiodes selected for our prototype are both optimized for $940$~nm infrared light, which rejects most ambient illumination at other frequencies. In order to further improve signal robustness, modulated illumination is used to reject any ambient light, even environmental $940$~nm light (i.e. sunlight). The device uses an analog front-end (ADPD103) to synchronously modulate the LEDs and sample the photodiodes. Each LED is illuminated for $3~\mu$~seconds every $24~\mu$~seconds. The photodiode response is bandpass filtered and synchronously integrated. The result is a photodiode signal sensitive to changes in reflected light from the LEDs, but not from other light sources.

To minimize the effect of direct reflections of the cornea, only one LED is illuminated at a time. In this implementation, only two LEDs are used in a particular session; each LED is associated with the four nearest photodiodes. For a single frame of data, the first LED pulses four times while the four nearest photodiodes are integrated and sampled. This process repeats for the second LED and remaining photodiodes. The overall data rate is determined by the number of LED pulses and delay between LEDs. In this prototype, the output data rate was configured to be $400$~Hz. The electrical current consumed by the LED is also configurable and determines the signal-to-noise ratio. In this prototype, it was set to $67$~mA. Note that this is instantaneous current through the LEDs. At this current setting, the overall power consumption of the analog front-end and LEDs is only $16$~mW.

\subsection{Gaze tracking LEDs}
Our second prototype consists of 6 LEDs per eye, with each LED functioning as both light sources and sensors, and an Arduino Nano microcontroller per eye for controlling those LEDs. We designed our second prototype to be a standalone platform to remove the need for a chin rest in our subjective experimentation, so we inserted a 2K resolution HDMI display (TopFoison) into an off-the-shelf mobile VR headset (Samsung Gear VR), in which the sensing electronics were housed. LEDs are placed around the VR headset's magnifier lenses.

The anodes of LEDs were attached to digital IO pins of a microcontroller, while their cathodes were attached to analog-to-digital converter (ADC) pins of the same microcontroller. Each time an LED is to be used in sensor mode, it follows the three steps described in Figure \ref{nvgaze:fig:reverse-led}. LEDs have a soft-coded mechanism that adjusts exposure times, $\Delta~t_{exp}$, on a per LED basis, so that saturation caused by varying light conditions can be avoided for each LED. The LEDs are sampled in a round-robin fashion, such that one LED  records measurements while the remaining LEDs serve as emitters. In contrast to NextGaze, we chose to use the remaining LEDs as emitters to minimize the sensing time of an LED, and in turn minimize system latency.

\begin{figure*}[ht!]
   \centering
   \includegraphics[width=\textwidth]{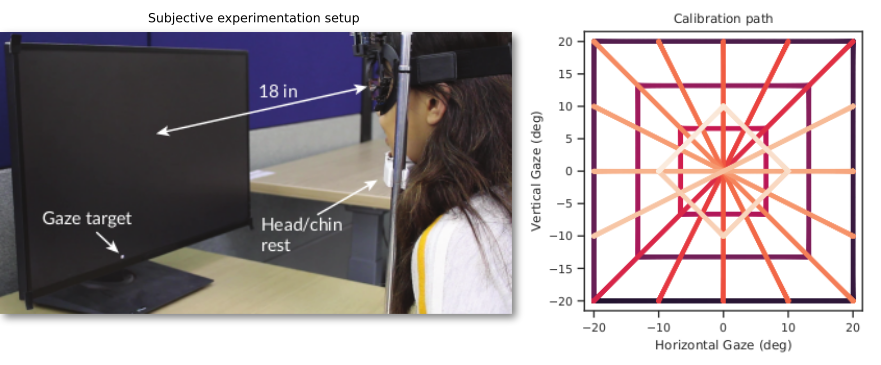}
   \caption{The calibration process requires a user to wear the prototype and follow a moving stimulus shown on an external computer screen with their gaze.
   Left: the user rests their chin in a head rest 18 inches from a fixed display. Right: the user is asked to follow a moving target that moves along multiple line segments with their gaze.}
   \label{fig:subjective_experimentation_setup_and_results}
\end{figure*}

The microcontroller communicates with the user interface application over a USB connection. This user interface application is developed using Kivy library~\cite{virbel2011kivy}. Our user interface application handles a number of tasks: (1) collecting measurements from each LED by requesting them from the two microcontrollers used for each eye, (2) updating the user interface, (3) estimating the gaze orientation, and (4) keeping logs related to captured data such as event timestamps, raw signals, estimations.

\section{Evaluation}
\label{section:evaluation}

\subsection{Evaluating NextGaze}

\subsubsection{Procedure}

We invited six participants (4 male, 2 female) to help us collect a dataset for evaluating the performance of NextGaze.
Due to NextGaze's design as a standalone device with no display, an external monitor was used to show the visual stimulus.
Participants that wore glasses were asked to remove them, however, contact lenses were allowed to be worn.  Participants were then asked to place their head in a desk mounted chin rest, putting his or her face 18 inches away from the monitor, as shown in Figure~\ref{fig:subjective_experimentation_setup_and_results} (left). A circular gaze target with a radius of 10 pixels (1.2 mm) is placed on the screen for calibration. The $x$ and $y$ coordinates of the target in screen-space can be computed from the screen distance and desired angular gaze coordinates.

Participants first engaged in a smooth pursuit task, following the guidelines outlined by Pfeuffer et al. \cite{pfeuffer2013pursuit}. The gaze target travels in a series of linear paths over $\pm 20\degree$ vertical and horizontal field of view as outlined in Figure~\ref{fig:subjective_experimentation_setup_and_results} (right). For each segment, the target smoothly accelerates from rest over 1 second, up to a max speed of $6\degree$ per second and then decelerates over 1 second back to rest at the end of the segment. This produces a dense sample of the gaze space. Particular care was taken to optimize the display pipeline to minimize any jitter of the target as it moved, as this could cause undesired saccades during the pursuit.

Following the smooth pursuit task, participants were then asked to visually fixate on 25 static targets displayed on a grid within $\pm 20 \degree$ horizontal and vertical. The saccadic nature of this task helped to diversify the dataset collected. The first twenty segments of the smooth pursuit data were used for training. The last four segments as well as the 25 fixation targets were used for testing.

For evaluation, we report error on a per-frame basis, taking the difference in angle between the target's position and the angle predicted by our modeling algorithm at every frame.  Although this protocol is limited by the strong assumption that the participant's gaze really is locked on the target, it is nonetheless standard procedure and we adopted it to produce results that can be compared with the related work.

\subsubsection{Modeling and Results}

First, a preprocessing step removes blinks from the signal by performing a Savitzky-Golay filter and removing areas where the filtered signal is different by more than a given threshold.
The remaining signal is then downsampled to 100 points along each line segment to avoid overfitting to points along the line.

\begin{figure}[]
   \centering
   \includegraphics[width=1.0\columnwidth]{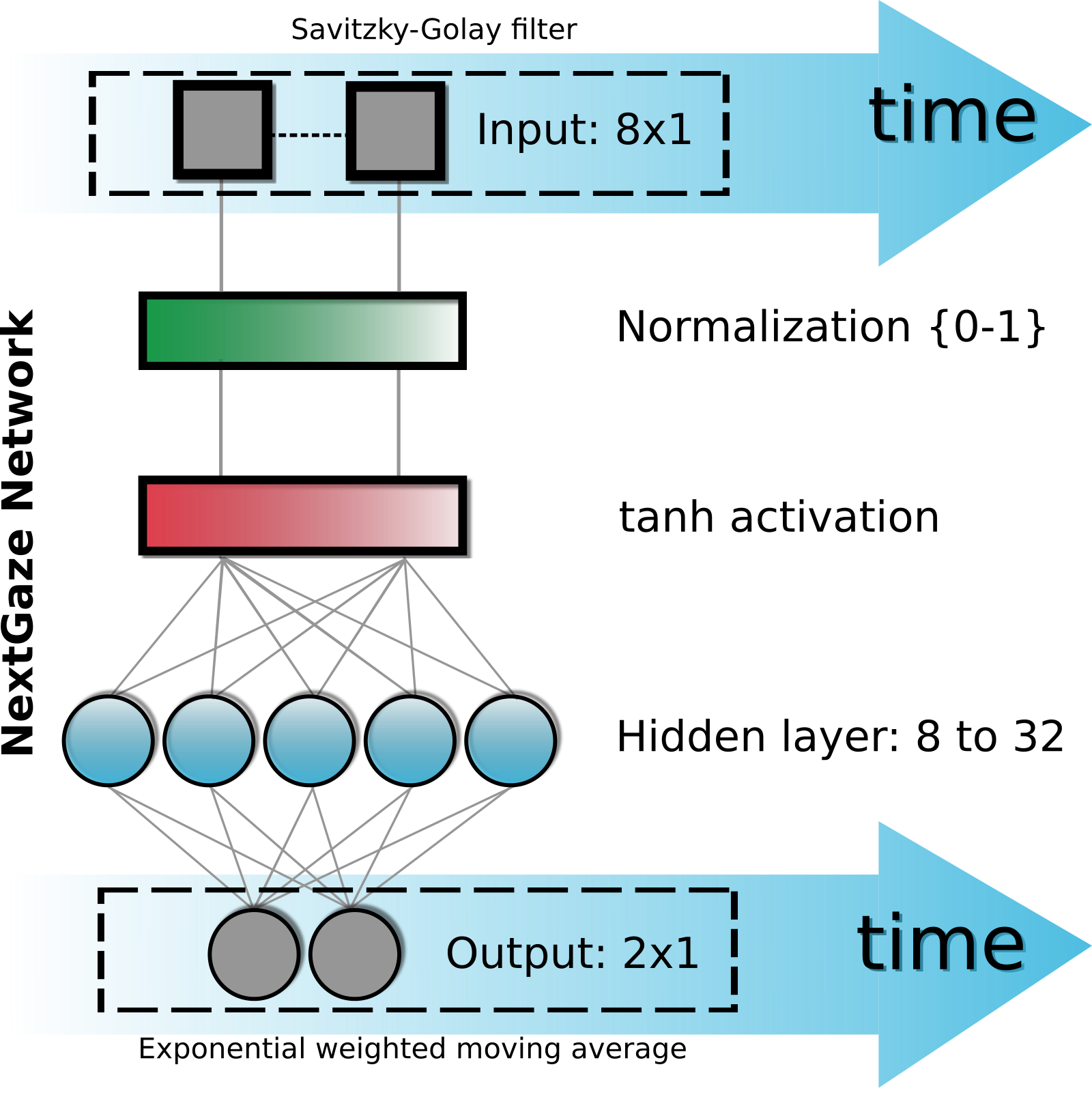}
   \caption{NextGaze Network. Measurements from photodiodes are passed through a normalization and activation step before going through a full connected neural network. Result is a two-dimensional output that represents gaze estimation along vertical and horizontal axes. To avoid temporal flickering or sudden variations in the results over the time, inputs of NextGaze network is filtered temporally using a Savitzky-Golay filter, while outputs of NextGaze network is temporally averaged using exponential weights.}
   \label{fig:next_gaze_network}
\end{figure}

The gaze model maps the photodiode output to gaze coordinates in degrees. Because the mapping is highly nonlinear, we leverage a neural network model to map the eight photodiode signals to 2D gaze. We first scale all photodiode outputs to 0 to 1 and then apply Principal Component Analysis (PCA) to reproject the data. The transformed signals are input to a network with $4$ hidden layers of $64$ nodes each and $\mathrm{tanh}$ activation functions. The model uses scikit-learn's implementation of a multi-layer perceptron neural network with a batch size of $4$, up to $500$ epochs, and an adaptive learning rate starting at $0.001$, as depicted in Figure \ref{fig:next_gaze_network}.

After calibration, the gaze is estimated and shown on screen in real time. Inferences from the model are post-processed using an exponential weighted moving average (EWMA) filter ($\alpha = 0.2$) to smooth the estimation results.
This filter was designed empirically to improve the real-time experience provided to the user.

\begin{table}[]
    \caption{Performance of the gaze model on different test sets. The model performs best on the smooth pursuit task.}
	\begin{tabularx}{\columnwidth}{r|X|X}
		\toprule
		\textbf{Task} & \textbf{Mean\newline Error} & \textbf{Standard \newline Deviation} \\
		
		\midrule
		
		Smooth Pursuit Validation &
		$1.68\degree$ &
		$0.56\degree$ \\
		
		Fixation Validation ($\pm20\degree$) &
		$2.67\degree$ &
		$0.98\degree$ \\

		Central Fixation Validation ($\pm10\degree$) &
		$2.35\degree$ &
		$0.58\degree$ \\
        
        \bottomrule
	\end{tabularx}
	\label{nvgaze:tab:res1}
\end{table}

In addition to using the real-time inferences for supporting the interactive system, we also ran an offline evaluation producing the results shown in Table \ref{nvgaze:tab:res1}. Mean error on the last four segments of the smooth pursuit task was $1.68\degree$. The error increased to $2.67\degree$ on the fixation task, but is slightly better when limiting the data to $\pm 10\degree$. Among the six trials, the best error on the $\pm 20\degree$ fixation task was $1.1\degree$ and the worst was $4.2\degree$.

\subsection{Evaluating LED2Gaze}

\subsubsection{Procedure}

We invited fourteen participants (10 male, 4 female) to help us collect a dataset for evaluating LED2Gaze.
Since LED2Gaze featured its own built-in display, no chin rest was necessary for this study.
Again, glasses were asked to be removed while contact lenses were allowed to be used.
Participants were seated in an office chair and asked to wear the headset.
Then they are engaged in a series of tasks similar to the procedure used in Study 1, again driven by a moving target.
First, participants performed a smooth pursuit task consisting of 9 segments.
The target guided the participant through two sessions of fixating on each of 16 targets forming a grid over the entire field of view of the headset ($101\degree$).  Finally, 66 random points were presented for the participant to fixate on.
In total, 9 segments of smooth pursuit and 98 saccadic fixation points were collected per participant.
For evaluation, the dataset was split such that the smooth pursuit, one session of 16 grid points, and 66 random points were used for training the model, and the second session of 16 grid were used for testing.

\subsubsection{Modeling and Results}

In our previous prototype, we employed a fully-connected neural network to address the nonlinearity of the mapping from photodiode readings to gaze orientation.  While the results were adequate, the complexity and abstract nature of neural networks can be tricky to interpret for humans and in turn difficult to improve the results of.  For our second prototype, we explored the use of a simpler, more interpretable approach, namely a GPR model.  Such models take the following general form:

$$
\begin{bmatrix} e_{x} \\ e_{y} \end{bmatrix} = k^{T}C^{-1} \begin{bmatrix} u_{x} \\ u_{y} \end{bmatrix}^{T}
$$

$$
k = \begin{bmatrix}
\kappa (s(t), \bar{c_{1}}) \\
... \\
\kappa (s(t), \bar{c_{p}})
\end{bmatrix}
$$

$$
C = \begin{bmatrix}
\kappa (\bar{c_{0}}, \bar{c_{0}}) & ... & \kappa (\bar{c_{0}}, \bar{c_{p}}) \\
... &  & \\ 
\kappa (\bar{c_{p}}, \bar{c_{0}}) & ... & \kappa (\bar{c_{p}}, \bar{c_{p}})
\end{bmatrix}
$$

Where $e_{x}$ and $e_{y}$ represents estimated gaze orientation along the $x$ and $y$ axes, respectively, $k^{T}$ represents a vector that contains the similarity measures between the captured $s(t)$, and the calibration vectors $\bar{c_{p}}$. Finally, $u_{x}$ and $u_{y}$ represent vectors that correspond to the $x$ and $y$ position of each $\bar{c_{p}}$.

Comparing a vector with another vector can be accomplished in multiple ways. In evaluating multiple different distance measures (Cosine, Minkowski, Manhattan, Canberra) \cite{bray1957ordination, lance1966computer, Rasmussen06gaussianprocesses}, we found that the Minkowski distance measure to be the most effective when used with the GPR algorithm.

\begin{table}[]
    \caption{LED2Gaze's performance on the fourteen participant dataset.}
	\begin{tabularx}{\columnwidth}{r|X|X|X}
		
		\toprule
		
		\textbf{Participant} &
		\textbf{Mean\newline Error} &
	    \textbf{Median\newline Error} &
	    \textbf{Standard \newline Deviation} \\
		
		\midrule
		
		Average &
		$1.57\degree$ &
		$1.12\degree$ &
		$2.00\degree$\\
        \bottomrule
	\end{tabularx}
	\label{nvgaze:tab:res2}
\end{table}

The results from this evaluation are shown in Table \ref{nvgaze:tab:res2}. Mean error was improved to $1.57\degree$ when compared to our previous prototype. Among the fourteen participants, the best per-participant error was $1.1\degree$ (equal to the best performance of NextGaze) and the worst was $2.1\degree$ (improving the worst performance of NextGaze by half).
Furthermore, we empirically found that the output was sufficiently smooth such that no post-processing such as the EWMA filter used by NextGaze was needed.


\section{Discussion}
\label{section:discussion}

\subsection{Towards Deploying Sparse Optical Gaze Trackers}

We discuss some practical considerations needed for spatially-sparse optical gaze trackers to be deployed in devices to be used in natural conditions.

\medskip

\noindent\textbf{User and Session Dependence.} In this work, our modeling and analysis reported per-session and per-user accuracy results.
On the other hand, in practice, gaze trackers are ideally session and user independent, such that the system can be put on and immediately used.
Further investigations are required to understand how much, if any, per-user or per-session calibration is required.
For example, it might be only necessary that the user follows a calibration procedure the very first time they use the system (i.e. user registration), or briefly calibrate the system with a few points (i.e. session registration).

\medskip
\noindent\textbf{Synthetic Data}
\label{synthetic_data}
Although our experiments only used real sensor data collected from participants for training and testing, there is an opportunity for making use of the synthetic data generated by the simulation framework to produce more robust models.
For example, the synthetic data could be used in a Bayesian framework for augmenting the signals recorded from participants, increasing the size of the training set \cite{tran2017bayesian}.
Alternatively, recent advances in neural network research have shown the potential for directly using synthetic data in the training procedure for more robust models. \cite{learning_from_simulated_and_unsupervised_images}

\medskip
\noindent\textbf{Wearable Accessories.} Factors such as prescription glasses, contact lenses, eye color, eyelashes, and mascara influence data quality~\cite{nystrom2013influence}. A common practice for avoiding usage of prescription glasses and contact lenses in VR/AR near-eye displays comes in the form of an add-on inset lens in the optical path of a near-eye display. On the other hand, next generation computational displays promises algorithmic approaches to the problem of eye prescriptions by using active components that can support various focus levels~\cite{chakravarthula2018focusar}.

In our studies, participants were asked to remove prescription glasses. However, there were subjects that wore contact lenses and mascara during experiments. Contact lenses are known to form air bubbles in between the cornea of an eye and a contact lens resulting in unintended dynamic reflections, and mascara can create a false glint, introducing noise into our sensor signals and causing robustness issues in camera based gaze tracking hardware~\cite{nystrom2013influence}.
Although our experiments did not reveal any particular robustness issues against wearing contact lenses or mascara, we also did not specifically control for it, and this issue remains not only an open research question but also a major challenge for deploying gaze tracking systems.

\medskip

\noindent\textbf{Compensating for Slippage.} In addition to per-session and per-user calibration, wearable devices also often face the additional challenge of within-session changes. Sparse optical gaze trackers are particularly sensitive to slippage of the device. Small millimeter-level shifts can cause significant changes in the observed sensor values. We conducted some initial experiments to explore the feasibility of compensating for slippage. We added four additional photodiodes in the device oriented toward the nose bridge and side of the face to collect signals corresponding to the position of the device on the face.
Preliminary results suggest that a monolithic model that incorporates both eye-tracking sensors and face-tracking sensors may be capable of gaze tracking that is invariant to sensor shifts.
Further investigation and additional training data is needed to demonstrate a full implementation.



\subsection{Opportunities in Gaze-Based Interfaces}

Gaze tracking has the potential to enable significant advances in interacting with mixed reality.
While gaze tracking research has traditionally optimized for greater tracking accuracy, we suggest that other facets of gaze tracking can be just as important to the user experience.
This subsection describes human perception and device interaction opportunities for gaze-based interfaces that span the spectrum of power, speed and latency, and accuracy requirements.

\subsubsection{Human Perception}


\noindent\textbf{Virtual Social Avatars.} Mixed reality offers the possibility of immersive telepresence through virtual avatars, or digital representations of oneself.
Literature in psychology has shown that the eyes convey a significant amount of information in the interaction between two people.
To improve the quality and immersiveness of telepresence interactions, gaze tracking is needed to drive the avatar \cite{garau2001impact}.
As social cues, it is important to achieve a high sample rate and low latency for the animated avatar to seem responsive \cite{nabiyouni2017relative, louis2019real}.
In addition, low power consumption is needed as the networking requirements of a video call already consume significant amounts of power.

\medskip

\noindent\textbf{Foveated Rendering.} Human eyes have maximum visual acuity in the fovea, a region in the retina of the eyes.
Areas outside of the fovea are perceived with less clarity.
Research in HMDs has explored the concept of “foveated rendering”, in which only the region the user is visually attending to is rendered with full quality, and has shown significant savings in computational requirements \cite{patney2016perceptually, guenter2012foveated}.
However, foveated rendering requires understanding the gaze direction of the eye to begin with.
As a technique for reducing computational requirements, it is natural that the sensing technique it relies on should similarly use low computation and low power.
Similar to animating avatars, the latency required of gaze tracking for foveated rendering needs to be low (less than 50 ms) \cite{albert2017latency}.

\subsubsection{Device Interaction}


\noindent\textbf{Activity Recognition.} Related work has shown that both camera-based and EOG-based gaze trackers can be used for recognizing activities of daily living, such as detecting reading and counting how many words have been read \cite{kai_reading}.
Such signals can be indicators of visual attention, passively disabling distractions such as notifications or automatically turning pages.
The use of EOG glasses in this related work exemplifies the notion that high accuracy is not needed to create a useful interface.

\medskip

\noindent\textbf{User Interfaces.} With clever design of an user interface, varying degrees of gaze tracking error can be useful and effective.
Bubble Gaze Cursor and Bubble Gaze Lens lessens the required gaze tracking accuracy by implementing an area cursor with a magnifying glass feature, essentially dynamically increasing the effective selection region \cite{choi2020bubble}.
Orbits further reduces the accuracy needed by presenting different moving visual stimuli, and simply confirming selection by measuring correlation between the gaze tracking results and the movements of the stimuli \cite{esteves2015orbits}.
A similar system, implemented using EOG glasses, show how the lower accuracy requirements can also alleviate the power consumption of a system with an effective interface \cite{dhuliawala2016smooth}.
Naugle and Hoskinson~\cite{naugle2013two} demonstrated that the coarsest gaze tracker, only recognizing whether a user is visually attending to a display or not, can be used to quickly interact with a head-mounted display while saving up to 75\% of the head-mounted display's normal power consumption. 




\subsection{Future Work}




\noindent\textbf{Exploring Multimodal Sensing.} In this paper, we have explored the use of photodiodes and reverse-driven LEDs for sensing gaze as an alternative to camera-based approaches.
While NextGaze featured two cameras, and videos were recorded, that data was never used for gaze inference.
In the future, we plan to explore how sensor fusion might help leverage the camera data in conjunction with the signals from our single-pixel detectors.
For example, the camera data might be used for periodic self-calibration, or as a higher accuracy fall-back when needed.
We are also interested in exploring how the single-pixel signals can be used to fill in the gaps between camera frames, such as a way of informing an interpolation function.

In addition to cameras, other sensors could potentially be used in tandem with our single-pixel detectors.
For example, strain gauges have been previously used to measure facial movements \cite{li2015facial}, and electrode arrays have been used to capture a variety of biosignals from the face \cite{bernal2018physiohmd}.

\medskip

\noindent\textbf{Beyond Gaze Tracking.} To make a compelling case of including a given sensor in future devices, such a technique should be able to serve multiple purposes.
There has been prior work using single-pixel detectors for facial action tracking and recognition~\cite{buccal} and for vital sign monitoring, such as heart-rate and blood oxygenation~\cite{cennini2010heart}.
We will explore such opportunities with our technique beyond gaze tracking in the future.

\section{Conclusion}
\label{section:conclusion}
In this paper, we explore the design space of gaze trackers that leverage sparse single-pixel optical sensing techniques.
Our rendering-based simulation framework enables accurate and rapid exploration of this design space. We present two wearable gaze tracking devices built on these techniques with designs grounded in insights gained from simulation results.
NextGaze explores the use of infrared LEDs and photodiodes to estimate gaze in a low-power wearable device, while LED2Gaze builds on these ideas and introduces a path to further form-factor improvements by leveraging LEDs as both emitters and sensors. These prototypes demonstrate the feasibility of using discrete optical elements to realize high-speed, low-power gaze tracking devices suitable for wearable use in virtual and augmented reality devices.
\bibliographystyle{abbrv-doi}

\bibliography{references}
\end{document}